\documentclass[twocolumn,aps,prb,showpacs]{revtex4}
\usepackage{makeidx}
\usepackage{amsmath}
\usepackage{amssymb}
\usepackage{graphicx}
\usepackage{times}
\usepackage{graphics}
\usepackage{dcolumn}
\usepackage{bm}
\def\mathbi#1{\textbf{\em #1}}

\begin{document}

\title{ \sffamily\bfseries\large Long-range correlated random field and random anisotropy
  $\mathbi{O(N)}$ models: A functional renormalization group study }
\author{ \sffamily\bfseries\normalsize Andrei A. Fedorenko$^{1}$ and
  Florian K\"uhnel$^{\,1,2}$ \smallskip}
\affiliation{ $^{1}$CNRS-Laboratoire de Physique Th{\'e}orique de l'Ecole Normale Sup{\'e}rieure,%
 24 rue Lhomond, 75231 Paris,  France \\
$^{2}$Fakult{\"a}t f{\"u}r Physik, Universit{\"a}t Bielefeld, Postfach 100131, D-33501
Bielefeld, Germany  }

\date{January 21, 2007}

\begin{abstract}
We study the long-distance behavior of the $O(N)$ model in the presence of
random fields and random anisotropies correlated as  $\sim1/x^{d-\sigma}$ for
large separation $x$ using the functional renormalization group.
We compute the fixed points and analyze their regions of stability within
a double $\varepsilon=d-4$ and  $\sigma$ expansion.
We find that the long-range disorder correlator remains analytic but generates
short-range disorder whose correlator develops the usual cusp.
This allows us to obtain the phase diagrams in $(d,\sigma,N)$ parameter space
and compute the critical exponents to first order in $\varepsilon$ and  $\sigma$.
We show that the standard renormalization group methods
with a finite number of couplings used in previous studies of systems
with long-range correlated random fields fail to
capture all critical properties.
We argue that our results may be relevant to the behavior of $^3$He-A
in aerogel.
\end{abstract}

\maketitle

\section{Introduction}
\label{sec1}

The effect of weak quenched disorder on large-scale properties and
phase diagrams of many-body systems attracted considerable
attention for decades. Despite significant efforts, there still
remain many open questions. The prominent example is the $O(N)$
model in which  an $N$-component order parameter (the
magnetization in the spin system notation) is coupled to a
symmetry-breaking random field. For $N=1$, it is known as the
random field Ising model (RFIM).\cite{nattermann98} For $N>1$, one
has to distinguish the  random field (RF) case where the order
parameter couples linearly to disorder, and the random anisotropy
(RA) case, where the coupling to disorder is bilinear. These models
are relevant for a variety of physical systems such as amorphous
magnets,\cite{harris73} diluted antiferromagnets in a uniform
external magnetic field,\cite{fishman79} liquid crystals in porous
media,\cite{LQ} critical fluids in aerogels,\cite{He3} nematic
elastomers,\cite{feldman01} and vortex phases of impure
superconductors. \cite{blatter94} For $N=1$, the RA model reduces
to the random-temperature model, where the randomness couples to
the local energy density as, for example, in diluted
ferromagnets.\cite{stichcombe-83} In contrast to the systems with
random-temperature-like disorder, the RF and RA models suffer from
the so-called dimensional reduction (DR). A straightforward analysis
of the Feynman diagrams giving the leading singularities
yields to all orders that the critical behavior of the RF $O(N)$ model in $d$
dimension is the same as that of the pure system in $d-2$
dimensions.\cite{young77} Consequently, the lower critical dimension is
$d^{\mathrm{DR}}_{\mathrm{lc}}(N=1)=3$ for Ising-like systems and
$d^{\mathrm{DR}}_{\mathrm{lc}}(N>1)=4$ for systems with continuous symmetry.
This can elegantly be demonstrated using supersymmetry.\cite{parisi79} However,
simple Imry-Ma arguments show that the lower critical dimension of
the RFIM is $d_{\mathrm{lc}}(N=1)=2$.\cite{imry75} The deviation
from DR is also confirmed by the high-temperature
expansion.\cite{gofman96} Thus DR breaks down, rendering
standard field theoretic methods useless.

Another known problem where the perturbation theory is spoiled by
DR is elastic manifolds in  disordered media. There, two
methods were developed to overcome difficulties related to DR:
the Gaussian variational approximation (GVA) in replica space  and
the functional renormalization group (FRG). The GVA  is supposed
to be exact in the limit $N\to\infty$. Unfortunately, this
approach when applied to the RF problem
leads to very complicated equations which do not allow us to
compute the critical exponents.\cite{mezard92} Considering the RF
$O(N)$ model, Fisher \cite{fisher85} showed that expansion around the lower
critical dimension $d_{\mathrm{lc}}=4$ generates an infinite number
of relevant operators which can be parametrized by a single
function. However, he found that the corresponding one-loop FRG equation
has no analytic fixed point (FP) solution.
Only recently, using the progress in the elastic manifold
problem,\cite{fisher86,nattermann92,chauve01,ledoussal02} it was realized that
the scaling properties of systems exhibiting metastability are encoded in
a nonanalytic FP. Feldman \cite{feldman02} has shown that, indeed, in
$d=4+\varepsilon$ and for $N\ge N_{\mathrm{c}}\approx 3$ (a more  precise
computation\cite{doussal06} gives $N_{\mathrm{c}}=2.834\,74$), there is a
nonanalytic FP with a cusp at the origin. This FP
provides the description for the ferromagnetic-paramagnetic phase
transition in the RF $O(N)$ model and allows one to compute the
critical exponents which are different from the DR prediction.
Recently, Le Doussal and Wiese \cite{doussal06} extended the FRG
analysis to two-loop order.
The extension beyond one-loop order is highly nontrivial due to
the nonanalytic character of the renormalized effective action,
which leads to anomalous terms in the FRG equation.\cite{ledoussal02}
The two-loop calculations were also independently performed in
Ref.~\onlinecite{tarjus06}, and the truncated exact FRG was proposed
in Ref.~\onlinecite{tarjus04}.
The more accurate analysis of the FRG
flows for the RF model showed that for $N>N^*=18+O(\varepsilon)$
there is a crossover to a FP with weaker nonanalyticity resulting
in the DR critical exponents.\cite{doussal06,tarjus06,tarjus04,sakamoto06}
A similar picture was found for the RA $O(N)$ model with the main difference
that $N_{\mathrm{c}}=9.4412$ and $N^*=\infty$.\cite{doussal06}

A more peculiar issue concerns the phase diagram of the RF and RA
models below $d_{\mathrm{lc}}$. It is known that for the RF model
and models with isotropic distributions of random anisotropies
 true long-range order is forbidden below $d_{\mathrm{lc}}=4$
(for anisotropic distributions,  long-range order can occur even
below $d_{\mathrm{lc}}$).\cite{dudka84} However,
quasi-long-range order (QLRO) with zero order parameter and an
infinite correlation length can persist even for
$d<d_{\mathrm{lc}}$. For instance, the GVA predicts that the
vortex lattice in disordered type-II superconductors can form the
so-called Bragg glass exhibiting slow logarithmic growth of
displacements.\cite{bragg} This system can be mapped onto the RF
$O(2)$ model, in which the Bragg glass corresponds to the QLRO
phase. Indeed, for $N <N_c$ and $d<d_{\mathrm{lc}}$, the FRG
equations have attractive FPs which describe the QLRO phases of RF
and RA models. \cite{feldman00} In order to study the transition
between the QLRO phase and the disordered phase expected
in the limit of strong disorder, one has to go beyond the one-loop
approximation.
The truncated exact FRG (Ref.~\onlinecite{tarjus04}) and the two-loop
FRG (Ref.~\onlinecite{doussal06}) performed using a double  expansion in
$\sqrt{|\varepsilon|}$ and $N-N_{\mathrm{c}}$ give access to a different
singly unstable FP which is expected to control the transition.
Both methods give qualitatively similar pictures of the FRG flows:
the critical and attractive FPs merge in some dimension
$d^*_{\mathrm{lc}}(N)<d_{\mathrm{lc}}$ that is therefore the lower
critical dimension of the QLRO-disordered transition.
For the RF $O(2)$ model, both methods predict approximately
the same lower critical dimension
$d^*_{\mathrm{lc}}\approx 3.8(1)$, and thus, suggest that there is
no Bragg glass phase in $d=3$. However, one has
to take caution when extrapolating results obtained for small
$\sqrt{|\varepsilon|}$ and $N-N_{\mathrm{c}}$.  Moreover, in contrast
to the model of Refs.~\onlinecite{feldman00} and \onlinecite{doussal06} which
belongs to the so-called ``hard-spin'' models,
the system studied in Ref.~\onlinecite{tarjus04} corresponds  to
``soft spins'', and thus, is expected to belong to a different
universality class. In terms of vortices, the soft spin RF model
allows for topological defects which destroy the Bragg glass.

Most studies of the RF and RA models are restricted  to either
short-range (SR) correlated disorder or uncorrelated pointlike
defects. However, real systems often contain long-range (LR)
correlated disorder or extended defects in the form of linear
dislocations, planar grain boundaries, three-dimensional cavities,
etc. Systems with anisotropic orientation of extended defects can
be described by a model in which all defects are strongly
correlated in $\varepsilon_d$ dimensions and randomly distributed
over the remaining $d-\varepsilon_d$ dimensions. The case
$\varepsilon_d=0$ is associated with uncorrelated pointlike
defects, while extended columnar or planar defects are related to
the cases $\varepsilon_d=1$ and 2, respectively. The critical
behavior of the $O(N)$ model with random-temperature-like extended
defects  was  studied in
Refs.~\onlinecite{dorogovtsev-80,boyanovsky-82,lawrie-84,fedorenko-04}
using a perturbative RG analysis in conjunction with a double
expansion in $\varepsilon=4-d$ and $\varepsilon_d$.

In the case of an isotropic distribution of disorder, power-law
correlations are the simplest example with the possibility for a
scaling behavior with new FPs and new critical
exponents. The critical behavior of systems with random-temperature
disorder correlated as  $1/x^{d-\sigma}$ for large
separation $x$ was studied in
Refs.~\onlinecite{weinrib-83,korucheva-98,fedorenko-00}. The power law
can be ascribed to extended defects of internal dimension $\sigma$,
distributed with random orientation in $d$ dimensional space.
In general, one would probably not expect a pure power-law decay of correlations.
However, if the correlations of defects arise from different sources with a
broad distribution of characteristic length scales, one can expect
that the resulting correlations will, over several decades, be
approximated by an effective power law.\cite{weinrib-83}
Power-law correlations with a noninteger
value $\sigma$ can be found in systems containing defects with
fractal dimension $\sigma$. \cite{yamazaki-88} For example, the
behavior of $^4\mathrm{He}$ in aerogels is argued to be described by
an XY model with LR correlated defects. \cite{vascuez-03} This is
closely related to the behavior of nematic liquid crystals enclosed
in a single pore of aerosil gel which was  recently studied in
Ref.~\onlinecite{feldman-04}, using the approximation in which the pore
hull is considered a disconnected fractal. The FRG was used to investigate
the statics and dynamics of elastic manifolds in media with LR correlated
disorder in Ref.~\onlinecite{fedorenko06}.
The critical behavior of the $O(N)$ model with LR correlated RF
was studied in Refs.~\onlinecite{kardar83,chang84,bray86}.
However, the methods used in these works fail to describe properly the case
of SR correlated RF. Therefore, there is a necessity to reexamine the critical
behavior of the LR RF model using methods which are successful in the SR case.

In the present paper, we study the  LR correlated RF and RA $O(N)$ models
using the FRG to one-loop order. The paper is organized as follows.
In Sec.~\ref{sec-model}, we introduce the LR RF and RA models, and derive the FRG equations.
In Sec.~\ref{sec-rf}, we study the LR RF model. In Sec.~\ref{sec-ra}, we consider
the LR RA model and discuss the application to superfluid $^3$He in aerogels.
The final section summarizes our results.


\section{Model and FRG equations}
\label{sec-model}

The large-scale behavior of the $O(N)$ symmetric spin systems at low temperatures
can be described by the nonlinear $\sigma$ model with the Hamiltonian
\begin{eqnarray} \label{H0}
 \mathcal{H}\left[ \vec{s}\, \right] = \int d^d x \left[
   \frac{1}{2} (\nabla \vec{s}\,)^2 + V(x,\vec{s}) \right],
\end{eqnarray}
where $\vec{s}( x )$ is  the $N$-component classical spin  with a fixed-length constraint
$\vec{s}\,^2=1$.  $\mathrm{V}(x,\vec{s})$ is the random disorder potential, which can be
expanded in spin variables as follows:
\begin{equation} \label{V0}
V(x,\vec{s})= \sum\limits_{\mu=1}^{\infty}\sum\limits_{i_1...i_ {\mu}}
-h^{(\mu)}_{i_1...i_ {\mu}}(x) s_{i_1}(x)...s_{i_{\mu}}(x).
\end{equation}
The corresponding coefficients have simple physical interpretation:
$h^{(1)}_i$ is a random field,
$h^{(2)}_{ij}$ is a random second-rank anisotropy, and $h^{(\mu)}$ are
general $\mu$th tensor anisotropies. As was shown in Ref.~\onlinecite{fisher85},
even if the system has only finite number of nonzero $h^{(\mu)}$, the
RG transformations generate an infinite set of high-rank
anisotropies preserving the symmetry with respect to
rotation $\vec{s}\to - \vec{s}$ if it is present in the bare model.
For instance, starting with only second-rank anisotropy corresponding to the RA model,
all even-rank anisotropies will be generated by the RG flow.
We will reserve the notation RA for the systems  which
have this symmetry and the notation RF for the systems which do not.
In the present work, we consider the case of Gaussian distributed long-range correlated
disorder with zero mean and cumulants given by
\begin{eqnarray}
\overline{h^{(\mu)}_{i_1...i_{\mu}}(x) h^{(\nu)}_{i_1...j_{\nu}}(x')}& =&
 \delta^{\mu\nu}\delta_{i_1 j_1}
...\delta_{i_{\mu} j_{\nu}} [r_1^{(\mu)}\delta(x-x') \nonumber \\
&&+ r_2^{(\mu)}g(x-x')],
\end{eqnarray}
with $g( x - x' ) \sim 1/|x-x'|^{d-\sigma}$. For the sake of convenience, we
fix the constant in Fourier space, taking $g(q)=1/q^{\sigma}$.

To average over disorder, we introduce $n$ replicas of the original system
and compute the replicated Hamiltonian
\begin{eqnarray}
\mathcal{H}_n &=&
   \int d^d x \left[
   \frac{1}{2} \sum_{a} (\nabla \vec{s}_a \,)^2
 - \frac{1}{2T} \sum_{a,b} R_1\big(\vec{s}_{a}( x )\cdot\vec{s}_{b}( x )\big) \right. \nonumber \\
&& - \left. \frac{1}{2T} \sum_{a,b}  \int d^d x'\, g( x - x' )\,
 R_2\big(\vec{s}_{a}( x )\cdot\vec{s}_{b}( x' )\big)
    \right],\ \ \ \ \    \label{Hn}
\end{eqnarray}
where $R_i(z)=\sum_{\mu}r_i^{(\mu)} z^{\mu}$. The properties of
the original disordered system  (\ref{H0}) and (\ref{V0}) can be
extracted in the limit $n\to 0$. According to the above definition of
the RF and RA models, the functions $R_i(z)$ are arbitrary in the
case of the RF and even for the RA. Power counting suggests that
$d_{\mathrm{lc}}=4+\sigma$ is the lower critical dimension for
both models.\cite{bray86}

At criticality  or in the QLRO phase, the correlation
functions of the order parameter exhibit scaling behavior.
In contrast to the models with temperature-like disorder  in the models
under consideration, the connected and disconnected
correlation functions may scale with different exponents.
This reflects metastability and the breaking of the DR.
For instance, the connected two-point function behaves as
\begin{equation} \label{cor-fun-con}
\overline{\langle \vec{s}(q)\cdot \vec{s}(-q) \rangle
 -   \langle \vec{s}(q)\rangle\cdot \langle\vec{s}(-q) \rangle } \sim q^{-2+\eta},
\end{equation}
while  the disconnected function scales as
\begin{equation} \label{cor-fun-dis}
\overline{\langle \vec{s}(q)\rangle \cdot \langle\vec{s}(-q) \rangle }
 -   \overline{\langle \vec{s}(q) \rangle }\cdot \overline{\langle \vec{s}(-q) \rangle}
  \sim q^{-4+\overline{\eta}}.
\end{equation}
Here, $\vec{s}(q)$ is the Fourier component of the order parameter and the
angle brackets stand for the thermal averaging.
Schwartz and Soffer\cite{schwartz85} proved that the exponents of the RF model obey
the inequality $2\eta\ge\bar{\eta}$.
Since in the RA case the coupling to disorder
is bilinear, the Schwartz-Soffer  inequality cannot be applied directly to
$\eta$ and $\bar{\eta}$. In the RA case, it is convenient to
introduce the correlation functions of the form Eqs.~(\ref{cor-fun-con}) and
(\ref{cor-fun-dis}) not for $\vec{s}$, but for the field
$m_i=[s_i]^2-1/N$, and define exponents $\eta_2$ and $\bar{\eta}_2$
which satisfy the Schwartz-Soffer type inequality:
$2\eta_2\ge\bar{\eta}_2$. \cite{feldman00} Vojta and Schreiber\cite{vojta95}
generalized this inequality to correlated RF and obtained
a more restrictive bound $2\eta-\sigma\ge\bar{\eta}$ since $\sigma>0$.
For RA models similar arguments lead to $2\eta_2-\sigma\ge\bar{\eta}_2$.

To derive the one-loop FRG equations, we straightforwardly
generalize the methods developed in
Refs.~\onlinecite{fisher85},
\onlinecite{feldman00} and \onlinecite{fedorenko06}
to model (\ref{Hn}). We
express the order parameter $\vec{s}_a$ as a
combination\cite{feldman00}
\begin{equation}
  \vec{s}_{a}(x) = \vec{n}_a(x)\sqrt{1-\vec{\pi}_a^{\,2}(x)} + \vec{\pi}_a(x)
\end{equation}
of a fast field $\vec{\pi}_a$ fluctuating at small scales
$\Lambda<q<\Lambda_0$
which is orthogonal to a slow field $\vec{n}_a$ of unit length,
changing at scales $q<\Lambda$.
Here, $\Lambda_0$ is the UV cutoff and $\Lambda\ll\Lambda_0$.
The field $\vec{n}_a$ can be considered as the coarse-grained
order parameter (local magnetization) whose fluctuations at low temperature
are small, $\langle\vec{\pi}_a^2 \rangle\ll1$.
Integrating out the fast variables $\vec{\pi}_a$,
we rescale in such a way that the effective Hamiltonian of the slow fields
$\vec{n}_a$ would have the structure of the bare Hamiltonian (\ref{Hn}).
It is convenient to change variable to $z=\cos\phi$.
The FRG equations  to first order in $\varepsilon$ and $\sigma$ are given by
\cite{florian-thesis}
\begin{subequations}
\begin{eqnarray}
    \partial_{\ell} R_1(\phi)
        &=& - \varepsilon R_1(\phi) + \frac12 \big[R_1''(\phi) + R_2''(\phi) \big]^2 - A R_1'' (\phi)\nonumber \\
        &&  - (N-2) \bigg\{ 2A R_1(\phi) + A R_1'(\phi)\cot\phi \nonumber \\
        &&  - \frac1{2\sin^2{\phi}} \big[ {R_1'(\phi) + R_2'(\phi) } \big]^2 \bigg\}, \label{frg-1} \\
    \partial_{\ell} R_2(\phi)
        &=& - (\varepsilon-\sigma) R_2(\phi) - \Big\{ (N-2)\big[2R_2(\phi) \nonumber \\
        &&  + R_2'(\phi) \cot\phi\, \big] +  R_2''(\phi)\Big\} A, \ \ \label{frg-2}
\end{eqnarray}
\end{subequations}
where $\partial_{\ell}:=-\partial/\partial \ln \Lambda$. We have absorbed the factor of $1/(8\pi^2)$
in redefinition of $R$ and introduced
\begin{equation}
  A = R_1''(0)+R_2''(0).
\end{equation}
In terms of the variable $\phi$, the functions $R_i(\phi)$
become periodic with period $2\pi$ in the  RF case and $\pi$ in the RA case.
The flow equation for the temperature to one-loop order reads
\begin{equation} \label{T}
    \partial_{\ell} \ln T = -(d-2) - (N-2) A.
\end{equation}
According to Eq.~(\ref{T}), the temperature is irrelevant for $d>2$ and sufficiently
small $A$. Although we expect $A=O(\varepsilon,\delta)$ in the vicinity of a FP,
one has to take caution whether the found FP survives in three dimensions.\cite{feldman00}
The scaling behavior of the system is controlled by a zero-temperature
FP of Eqs.~(\ref{frg-1}) and (\ref{frg-2}) $[R_1^*,R_2^*,A^*]$, such that
$\partial_{\ell} R_i^*=0$.
An attractive FP
describes a phase, while a singly (unidirectionally) unstable FP describes
the critical behavior.
The critical exponents are determined by the FRG flow in the vicinity of the FP and
to one-loop order are given by
\begin{subequations}
\begin{alignat}{2} \label{exp-eta}
    \eta
        &=  - A^*\;,&           \bar{\eta}      &=  -\varepsilon - (N-1)A^*,\\
    \eta_2
        &=  -(N+2) A^*\;,& \quad   \bar{\eta}_2 &=  -\varepsilon - 2 N A^*.
\end{alignat}
\end{subequations}
It is convenient to introduce the following reduced variables:
\begin{subequations}
\begin{eqnarray}
  r_i(\phi)&=&R_i(\phi)/(\varepsilon-\sigma), \\
  a&=&A/(\varepsilon-\sigma), \\
  \hat{\varepsilon}&=&\varepsilon/(\varepsilon-\sigma).
\end{eqnarray}
\end{subequations}
To check the stability of the FP $[r_1^*,r_2^*,a]$, we linearize  the flow equations
around this FP: $r_i(\phi)=r_i^*(\phi)+y_i(\phi)$ and obtain
\begin{subequations}
\begin{eqnarray}
    \lambda y_1(\phi)
        &=&     -\hat{\varepsilon} y_1(\phi) + \big[r_1^{*\prime\prime}(\phi) + r_2^{*\prime\prime}(\phi) \big]
                \big[y_1''(\phi) + y_2''(\phi) \big]\nonumber \\
        &&      - a y_1''(\phi) - a_0 r_1^{*\prime\prime}(\phi)
                -(N-2) \Big\{ 2a_0 r_1^*(\phi)  \nonumber \\
        &&     +  a_0 r_1^{*\prime}(\phi)\cot\phi + 2a y_1(\phi) + a y_1'(\phi)\cot\phi   \nonumber \\
        &&      - \big[ r_1^{*\prime}(\phi) + r_2^{*\prime}(\phi)  \big] \nonumber \\
        && \times
                        \big[ y_1'(\phi) + y_2'(\phi)  \big]/\sin^2{\phi} \Big\} , \ \ \ \ \ \ \ \ \
                        \label{frgl-1} \\
     \lambda y_2(\phi)
        &=&     - y_2(\phi) - a_0 \Big\{ (N-2)\big[2r_2^*(\phi) + r_2^{*\prime}(\phi) \cot\phi \big] \nonumber \\
        &&      + r_2^{*\prime\prime}(\phi) \Big\} - a \Big\{ (N-2)\big[2y_2(\phi) + y_2'(\phi) \cot\phi \big] \nonumber \\
        &&      + y_2''(\phi) \Big\}, \label{frgl-2}
\end{eqnarray}
\end{subequations}
where we have introduced the eigenvalues $\lambda$ (measured in units of $\varepsilon-\sigma$)
and defined $a_0=y_1''(0)+y_2''(0)$.
The FP is attractive if all $\lambda_i$ fulfill the inequality  $(\varepsilon-\sigma)\lambda_i<0$.
A singly unstable FP has only one eigenvalue $\lambda_1$ such
that $(\varepsilon-\sigma)\lambda_1>0$, which determines the third independent exponent
\begin{equation} \label{nu-def}
  \nu = 1/[\lambda_1(\varepsilon-\sigma)].
\end{equation}
This exponent characterizes the divergence of the correlation length
in the vicinity of transition.

\begin{table}[tbp]
\squeezetable \caption{LR RF model above the lower critical
dimension $d_{\mathrm{lc}}=4+\sigma$. The FP values of
$r_1^{*\prime\prime}(0)$,  $r_1^{*\prime\prime}(\pi)$ and the
relevant eigenvalue $\lambda_1$ computed numerically for different
$N$ and $\hat{\varepsilon}$. The last column is the relevant
eigenvalue $\lambda_1^{\mathrm{T}}$ obtained from the truncated
RG scheme of Ref.~\onlinecite{chang84} and computed using
Eq.~(\ref{nu-dr}).}
\label{tab-rf}%
\begin{ruledtabular}
\begin{tabular}{cdddddd}
\mbox{$N$} & \mbox{$\hat{\varepsilon}$}&  \mbox{$r_1^{*\prime\prime}(0)$}
 & \mbox{$ r_1^{*\prime\prime}(\pi)$}  & \mbox{$\lambda_1$}
 &   \mbox{$\lambda_1^{\mathrm{T}}$} \\ \hline
$4$        & 1.271 &    $-$1.0000     &      $ $0.5811   &  $ $1.271  &  $ $2.429      \\
           & 2     &    $-$0.4657     &      $ $0.0891   &  $ $1.218  &  $ $2.000      \\
           & 3     &    $-$0.3320     &      $ $0.0041   &  $ $1.198  &  $ $1.618      \\
           & 4     &    $-$0.2668     &      $-$0.0207   &  $ $1.192  &  $ $1.414      \\ \hline
$5$        & 2     &    $-$0.1743     &      $ $0.0192   &  $ $1.167  &  $ $1.366      \\
           & 3     &    $-$0.1132     &      $-$0.0073   &  $ $1.160  &  $ $1.225      \\
           & 4     &    $-$0.0819     &      $-$0.0138   &  $ $1.144  &  $ $1.158      \\ \hline
$6$        & 2     &    $-$0.0941     &      $ $0.0080   &  $ $1.145  &  $ $1.215      \\
           & 3     &    $-$0.0549     &      $-$0.0058   &  $ $1.127  &  $ $1.135      \\
\end{tabular}
\end{ruledtabular}
\end{table}

\section{Long-range Random field $O(N)$ model}
\label{sec-rf} We now focus on the phase diagram and critical
behavior of the LR RF $O(N)$ model. Kardar \textit{et
al.}~\cite{kardar83} studied the critical behavior of the LR RF
$O(N)$ model using a $\bar{\varepsilon}=d_{\mathrm{uc}}-d$
expansion around the upper critical dimension
$d_{\mathrm{uc}}=6+\sigma$. They found that,  to lowest order in
$\bar{\varepsilon}$, the critical properties are that of a pure
system in $d-2-\sigma$ dimensions; however, this generalized DR
was found to fail at higher orders. Chang and
Abrahams\cite{chang84}  applied to the LR RF $O(N)$ model a
low-temperature version of the RG.  Expanding around the lower critical
dimension $d_{\mathrm{lc}}=4+\sigma$, they obtained the RG
recursion relations for the three parameters $T$,
$\Delta_1$, and $\Delta_2$,  which in our notation correspond to
$\Delta_i=-R_i''(0)$. As we have shown in the previous section,
the truncated RG neglects an infinite number of relevant operators.
They found the nontrivial zero-temperature FP, which  in our
notation $\delta_i=\Delta_i/(\varepsilon-\sigma)$ reads
\begin{eqnarray}
 \!\!\! \delta_1^*&=&\frac{1}{(\hat{\varepsilon}-1)(N-3)^2},\ \ \ \
  \delta_2^*=\frac{\hat{\varepsilon}(N-3)-N+2}{(\hat{\varepsilon}-1)(N-3)^2}.\ \ \
    \label{delta-1-2}
\end{eqnarray}
The correlation length exponent $\nu$ is determined by Eq.~(\ref{nu-def}),
with the relevant eigenvalue $\lambda_1$ given by
\begin{eqnarray}
    \lambda^{\mathrm{T}}_{\mathrm{1}}
        &=& \frac{N-2}{N-3} - \frac{\hat{\varepsilon}}{2}
+\frac{\hat{\varepsilon}}{2}
\sqrt{1+\frac{4[N-2-\hat{\varepsilon}(N-3)]}{\hat{\varepsilon}\,^2(N-3)^2}}.
\ \ \ \ \ \ \label{nu-dr}
\end{eqnarray}
Note that the expression for the relevant eigenvalue reported in
Ref.~\onlinecite{chang84} is incorrect and gives values
which are several times larger than that computed using Eq.~(\ref{nu-dr}).
Although the truncated RG scheme when applied to the model with only SR correlated RF
($\Delta_2=0$) results in the DR FP $\Delta_{1\mathrm{DR}}^*={\varepsilon}/(N-2)$ and
exponent $\nu_{\mathrm{DR}}=1/{\varepsilon}$, the exponent~(\ref{nu-dr})
differs from the generalized DR prediction $\nu_{\mathrm{DR}}=1/({\varepsilon}-\sigma)$.
Nevertheless, one may doubt about the applicability of the truncated RG to the
LR RF $O(N)$ model even if the DR is broken.

\begin{figure}[tbp]
\includegraphics[clip,width=3.2in]{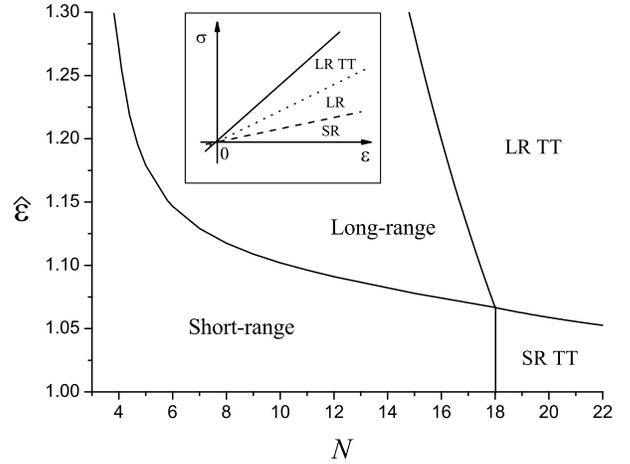}
\caption{The  stability regions of various FPs
corresponding to different patterns of the critical behavior above
the lower critical dimension, ${\varepsilon}>\sigma$. The borderline
between the SR and LR FPs is computed numerically using Eq.~(\ref{criterion-rf});
the borderline between the SR TT and LR TT regions is given by Eq.~(\ref{TT0})
and that between the LR and LR TT regions by Eq.~(\ref{TT}).
Inset: Schematic phase diagram on the $(\varepsilon,\sigma)$
plane for a particular value of $N\in(3,18)$. The solid line is
$\sigma=\varepsilon$; the dashed and dotted lines are given by
Eqs.~(\ref{criterion-rf}) and (\ref{TT}),  respectively.}
 \label{fig1-rf}
\end{figure}

We now reexamine the long-distance behavior of the LR RF $O(N)$ model by means of  the
full one-loop FRG derived in Sec.~\ref{sec-model}.
Equation~(\ref{frg-2}) is linear in the function $r_2(\phi)$ and can be solved analytically.
The FP solution fulfilling the RF boundary condition, i.e., $2\pi$ periodicity, is given by
\begin{equation}
  r_2^*(\phi)=-r_2^{*\prime\prime}(0)\cos\phi,
\end{equation}
which is an analytic function. Note that the analyticity of the LR part of the disorder
correlator was also revealed in the elastic manifold problem.\cite{fedorenko06}
This also gives us the FP value of $a$,
\begin{equation}
  a^*_{\mathrm{LRRF}}= - \frac{1}{N-3}, \label{A-rf}
\end{equation}
which, following Eqs.~(\ref{exp-eta}), completely fixes the values of the
critical exponents $\eta$ and $\bar{\eta}$,
\begin{eqnarray} \label{rf-exp}
  \eta_{\mathrm{LR}} = \frac{\varepsilon-\sigma}{N-3}, \  \  \  \  \
  \bar{\eta}_{\mathrm{LR}} = \frac{2\varepsilon-(N-1)\sigma}{N-3}.
\end{eqnarray}
Exponents (\ref{rf-exp}) satisfy the generalized Schwartz-Soffer inequality
at equality. This is at variance with the SR RF models, where the Schwartz-Soffer inequality
was found to be strict,\cite{feldman02}
but in agreement with the results for the  LR RF spherical model with long-range
interactions.\cite{vojta95}
One may conjecture  that the generalized Schwartz-Soffer inequality  is satisfied as
equality for any $N$ including the LR RF Ising model,\cite{vojta95}
but there is no argument that this persists in higher orders of loop expansion.

Let us first discuss the critical behavior of the LR RF model
above the lower critical dimension, $\varepsilon>\sigma$. Note
that $-a^*_{\mathrm{LRRF}}$  exactly coincides with
$\delta_1^*+\delta_2^*$ given by Eqs.~(\ref{delta-1-2}). Therefore
the truncated RG  would give the same values of $\eta$ and
$\bar{\eta}$ as the full FRG, at least to first order in
$\varepsilon$ and $\sigma$, if these exponents would be computed in
Ref.~\onlinecite{chang84}. To obtain the function $r_1^*(\phi)$ and
the amplitude $r_2^{*\prime\prime}(0)$, we integrate
Eq.~(\ref{frg-1}) numerically. Since the coefficients of
Eq.~(\ref{frg-1}) are singular at $\phi=0$ and $\pi$, to compute
the solution in the vicinity of these points, we use expansions of
$r_1^*(\phi)$ in powers of $|\phi|$ and $(\pi-\phi)^2$,
respectively. The expansion around $0$ is completely determined by
the value of $r_1^{*\prime\prime}(0)$,
\begin{eqnarray}
  r_1^*(\phi)& = &\frac{ 2(N^2-4N+3) r_1^{*\prime\prime}(0) + N-1}{2(N-3)
   [(N-3)\hat{\varepsilon} - 2(N-2)]} + \frac{r_1^{*\prime\prime}(0) \phi^2}{2}  \nonumber \\
  &\pm& \frac{\sqrt{r_1^{*\prime\prime}(0)(\hat{\varepsilon}-1)(N-3)^2+1}}{ 3(N-3)\sqrt{N+2}}
     |\phi|^3  +   O(\phi^4), \ \ \ \ \label{series1}
\end{eqnarray}
while to get the explicit expansion around $\pi$ we need know $r_1^{*\prime\prime}(0)$
and $r_1^{*\prime\prime}(\pi)$:
\begin{eqnarray}
&&  \!\!\!\!\!\!\!\! r_1^*(\phi)  =  (N-1) \nonumber \\
&&  \times \frac{\Big\{\big[r_1^{*\prime\prime}(0) +r_1^{*\prime\prime}(\pi)\big](N-3)+1\Big\}^2
  +2 r_1^{*\prime\prime}(\pi)(N-3)}{2(N-3)
   [(N-3)\hat{\varepsilon} - 2(N-2)]}   \nonumber \\
    &&+ \frac{r_1^{*\prime\prime}(\pi) (\pi-\phi)^2}{2} + O[(\pi-\phi)^4]. \label{series2}
\end{eqnarray}
From Eq.~(\ref{series1}), we see that the SR part of the disorder correlator is nonanalytic
at small $\phi$, so that we have to distinguish the left and right derivatives. In what
follows, we adopt $r_1^{(n)}(0)\equiv r_1^{(n)}(0^+) $.
We use numerical integration to continue the solutions given by expansions
(\ref{series1}) and (\ref{series2}) inside the interval $[0,\pi]$ and
match them by adjusting the shooting parameters $r_1^{*\prime\prime}(0)$
and $r_1^{*\prime\prime}(\pi)$.
Only the series with ``$+$'' in
Eq.~(\ref{series1}) can be matched with the solution computed using expansion (\ref{series2}).
Following the third term of  Eq.~(\ref{series1}),
the FP solution exists only if $-r_1^{*\prime\prime}(0)\le  \delta^*_1$.
We found that this inequality is always strict,
however, the difference $r_1^{*\prime\prime}(0)-(-\delta^*_1)$
becomes smaller in the limit of large $N$.
Thus the truncated RG fails to give the correct values of $r_1^{*\prime\prime}(0)$
and $r_2^{*\prime\prime}(0)$ although it gives the correct sum. Consequently,
the relevant eigenvalue $\lambda_1$ and the exponent $\nu$ obtained from
the truncated RG are also expected to be incorrect.
The computed values of $r_1^{*\prime\prime}(0)$ and
$r_1^{*\prime\prime}(\pi)$ are shown in Table~\ref{tab-rf}.

Let us check the stability of FPs. First, we examine the stability of the
SR FP $[r_{\mathrm{SR}}^*,r_2=0, a_{\mathrm{SR}}^*=r_{\mathrm{SR}}^{*\prime\prime}(0)]$
found numerically  in Ref.~\onlinecite{feldman00}.
Linearized about the FP, Eq.~(\ref{frgl-2}) can be solved
analytically giving $y_2=\cos \phi$ and $\lambda=-1-a_{\mathrm{SR}}^*(N-3)$.
The SR FP is stable against the introduction of LR disorder
if $(\varepsilon-\sigma)\lambda<0$.
Taking into account that
$\eta_{\mathrm{SR}}=-(\varepsilon-\sigma)a_{\mathrm{SR}}^*$
and $\bar{\eta}_{\mathrm{SR}}=-\varepsilon-(N-1)(\varepsilon-\sigma)a_{\mathrm{SR}}^*$, we
can rewrite the criterion of the SR FP stability in the following form
\begin{equation} \label{criterion-rf}
  \sigma < 2\eta_{\mathrm{SR}}-\bar{\eta}_{\mathrm{SR}},
\end{equation}
which was derived using general RG scaling arguments
in Ref.~\onlinecite{bray86}.  The regions of stability of the
LR and SR FPs are shown in Fig.~\ref{fig1-rf}.
Tissier  and  Tarjus (TT) argued that
for $N>18$, the cuspy SR FP becomes more than once unstable,
but a different critical (singly unstable) SR TT FP arises with
the function $R_1^*(\phi)$ being only $p\sim N$ times differentiable at
the origin.\cite{tarjus06}
Although this weak nonanalyticity should reflect some metastability in the
system,\cite{doussal06} it leads to the DR critical exponents. According to
Eq.~(\ref{criterion-rf}), the SR TT FP is stable with respect to the LR correlated
disorder for $N>18$ and
\begin{equation} \label{TT0}
  1<\hat{\varepsilon}<(N-2)/(N-3).
\end{equation}

\begin{figure}[tbp]
\includegraphics[clip,width=3.2in]{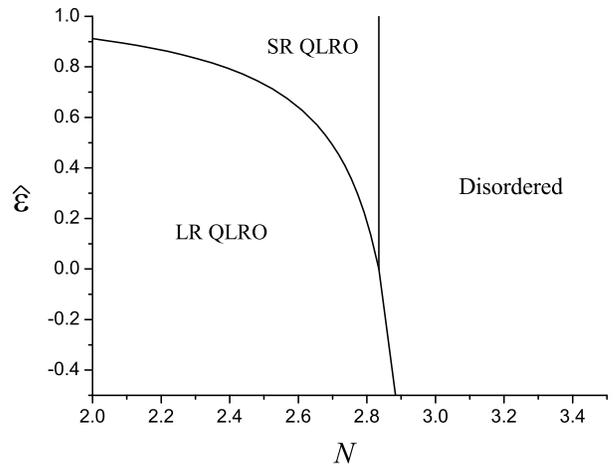}
\caption{Phase diagram of the RF model below the lower critical
dimension, $d<4+\sigma$ ($\varepsilon<\sigma$). } \label{fig2-rf}
\end{figure}

We now check the stability of the LR RF FP
$[r_1^*(\phi),\ r_2^*(\phi),\ a=-1/(N-3)]$.
Substituting the LR FP in Eqs.~(\ref{frgl-1}) and (\ref{frgl-2}),  we
obtain $y_2(\phi)=\cos \phi$ and
\begin{equation}
  a_0=-\frac{\lambda}{1+ r_1^{*\prime\prime}(0) (N-3)}.
\end{equation}
To compute the eigenfunction $y_1(\phi)$ and eigenvalue $\lambda$, we
solve Eq.~(\ref{frgl-1}) numerically using shooting and imposing
the $2\pi$-periodic boundary condition.
The obtained values of $\lambda_1$ are shown in Table \ref{tab-rf}.
As one can see from the table, the relevant eigenvalues $\lambda_1$ computed using the full
FRG, and consequently, the exponent $\nu$ differ significantly from that obtained
using the truncated RG. The critical exponents are expected to be continuous functions
of $\varepsilon$, $\sigma $, and $N$.
In the region controlled by the SR FP the correlation length exponent is
$\nu_{\mathrm{SR}}=1/\varepsilon$, independent on $N$.\cite{tarjus06}
The first line in Table~\ref{tab-rf}
corresponds to a point on the borderline separating the regions of stability
(see Fig.~\ref{fig1-rf}), and thus, on the borderline we indeed have
${\nu_{\mathrm{SR}}}={\nu_{\mathrm{LR}}}$.
We checked that the second eigenvalue $\lambda_2$ computed at the LR FP
vanishes exactly on the borderline,
indicating that the LR FP becomes twice unstable in the region dominated by the
SR FP. We also looked for an analog of the TT phenomena in the presence
of LR correlated disorder and found that for
\begin{equation} \label{TT}
  \hat{\varepsilon}>\frac{2\sqrt{N+7}+6}{N-3}
\end{equation}
there is a crossover to a new singly unstable LR FP of a TT type with
the function $R_1^*(\phi)$ being only $p\sim N$ times differentiable at
the origin and $R_2^*(\phi)\sim \cos\phi$. The values
of $R_1^{\prime\prime}(0)$ and $R_2^{\prime\prime}(0)$  at the
LR TT FP are simply given by Eq.~(\ref{delta-1-2}). Thus,
the LR TT FP leads to the critical exponents (\ref{rf-exp}) and
the relevant eigenvalue (\ref{nu-dr}) obtained from the truncated RG.

Finally, we discuss the  LR RF $O(N)$ model below the lower
critical dimension,  $\epsilon<\sigma$. For $N<3$, the FRG
equations have two attractive  FPs: the SR and LR,  which describe
the SR QLRO and LR QLRO phases, respectively. The phase diagram
computed from the stability analysis of different FPs is depicted in
Fig.~\ref{fig2-rf}. The exponents characterizing the power-law
behavior of the connected and disconnected correlation functions
in the LR QLRO phase are given by Eq.~(\ref{rf-exp}). The
physically interesting case $N=2$ describing  the Bragg glass
phase in the presence of the LR correlated disorder was considered
by one of the authors in Ref.~\onlinecite{fedorenko06}. In the
Bragg glass, the displacements $u(x)$ of a periodic structure (e.g.,
vortex lattice) grow logarithmically as
$\overline{(u(x)-u(0))^2}=\mathcal{A}_d\ln|x|$. Note that in
Ref.~\onlinecite{fedorenko06} the period was fixed to $1$ while in
the present work it is $2\pi$, so that the relation between the
universal amplitude $\mathcal{A}_d$ and the exponent $\eta$ is
given by $\eta=2\pi^2 \mathcal{A}_d$. For $N=2$, the crossover from
SR QLRO with the critical exponents
$\eta_{\mathrm{SR}}=|\varepsilon|\pi^2/9$ and
$\bar{\eta}_{\mathrm{SR}}=|\varepsilon|(1+\pi^2/9)$ to  LR
QLRO with the exponents $\eta_{\mathrm{LR}}=|\varepsilon|+\sigma$
and $\bar{\eta}_{\mathrm{LR}}=2|\varepsilon|+\sigma$ happens for
$\hat{\varepsilon} < 9/\pi^2$ (see Fig.~\ref{fig2-rf}).

\section{Long-range Random anisotropy  $O(N)$ model}
\label{sec-ra}

In this section, we study the LR RA case which corresponds to
$\pi$-periodic functions $r_i(\phi)$.  Equation~(\ref{frg-2}) has a
family of (at most) $\pi$-periodic solutions which can be expressed as
polynomials in $\cos\phi$ of power $p$ with
\begin{eqnarray}
  a^*_p=\frac1{4p^2+2(p-1)(N-2)}, \hspace{8mm} p=1,2,....
\end{eqnarray}
For instance, for $p=1$, we have
\begin{eqnarray}
  r_2^*(\phi)&=& \frac1{8N}[4r_1^{*\prime\prime}(0)-1][N\cos^2\phi-1],\label{r2-ra} \\
  a^*_{\mathrm{LRRA}}&=&\frac14, \label{a-ra}
\end{eqnarray}
and for $p=2$, we have $a_2^*=1/(2N+12)$ and
\begin{eqnarray}
  r^{(2)}_{2}&=&\frac{2(N+6)r_{1}^{(2)\prime\prime}(0)-1}{8(N+1)(N+2)(N+6)} \nonumber \\
  &\times& [3-6(N+2)\cos^2\phi+(N+2)(N+4)\cos^4\phi]. \nonumber
\end{eqnarray}
The stability analysis shows that due to the
inequality $a_p^*<a_1^*\equiv a^*_{\mathrm{LRRA}}$ for $p\ge2$ and $N\ge2$,
all FPs with $p>2$ are
unstable. It can be easily seen for $N=2$ when the functions $r^{(p)}_2$ ($p\ge 2$) become
$(\pi/p)$-periodic, and thus, are unstable with respect to a $\pi$-periodic perturbation.
Henceforth we consider only the LR FP determined by Eqs.~(\ref{r2-ra}) and (\ref{a-ra}),
which give the following values of the critical exponents
\begin{eqnarray}
  \eta_{\mathrm{LR}} &=& \frac{\sigma-\varepsilon}{4},\ \ \ \ \ \
   \bar{\eta}_{\mathrm{LR}} = \frac{\sigma}4(N-1) -\frac{\varepsilon}4(N+3),\ \ \
   \label{exp-ra-1} \\
  \eta_{2\mathrm{LR}} &=& \frac{(\sigma-\varepsilon)(N+2)}{4}, \ \
  \bar{\eta}_{2\mathrm{LR}} = -\varepsilon + \frac{N}2 (\sigma-\varepsilon).\ \ \ \
   \label{exp-ra-2}
\end{eqnarray}
Exponents (\ref{exp-ra-2}) satisfy the generalized Schwartz-Soffer inequality
$2\eta_2-\sigma\ge\bar{\eta}_2$ also at equality.
Analogously to the LR RF model, in order to obtain the function $r_1^*(\phi)$,  we
solve Eq.~(\ref{frg-1}) numerically using shooting and imposing the $\pi$-periodic
boundary condition. Since the coefficients of Eq.~(\ref{frg-1})
are singular at $\phi=0$,
we use an expansion of $r_1^*(\phi)$ in powers of $|\phi|$ which reads
\begin{eqnarray}
  r_1^*(\phi)& = &\frac{ (N-1)[1-8 r_1^{*\prime\prime}(0)]}{16(N-2+2
   \hat{\varepsilon})} + \frac{r_1^{*\prime\prime}(0) \phi^2}{2}  \nonumber \\
  &&\pm \sqrt{1+\frac{16r_1^{*\prime\prime}(0)(\hat{\varepsilon}-1)}{N+2}}
    \frac{ |\phi|^3}{12}  +   O(\phi^4). \ \ \ \ \label{series-ra}
\end{eqnarray}
As one can see from Eq.~(\ref{series-ra}), the SR disorder correlator
$r_1^*(\phi)$ has a cusp at the origin.
Only the solution with ``+'' in Eq.~(\ref{series-ra}) fulfills all conditions.

We now check the stability of the SR RA FP with respect to the
LR correlated disorder.
Analogously to the RF case, linearized around the FP, Eq.~(\ref{frgl-2})
allows for an analytical solution giving
$y_2=N\cos^2 \phi-1$ and $\lambda=-1+4a_{\mathrm{SR}}$.
The SR FP is stable if
$(\varepsilon-\sigma)\lambda<0$, which can be rewritten as
\begin{equation}
  \sigma < 2\eta_{2\mathrm{SR}}-\bar{\eta}_{2\mathrm{SR}}. \label{ra-ineq}
\end{equation}
Above the  lower critical dimension, $\varepsilon>\sigma$,
inequality (\ref{ra-ineq}) holds for all $N>N_c=9.4412$ so that
the SR FP is stable with respect to the weak LR correlated
disorder. Although the LR correlated disorder does not change the critical behavior
for $N>N_c$, it can modify the critical behavior which should exist
for $N<N_c$ but which is not accessible in the one-loop
approximation.\cite{feldman02,doussal06}

\begin{figure}[tbp]
\includegraphics[clip,width=3.1in]{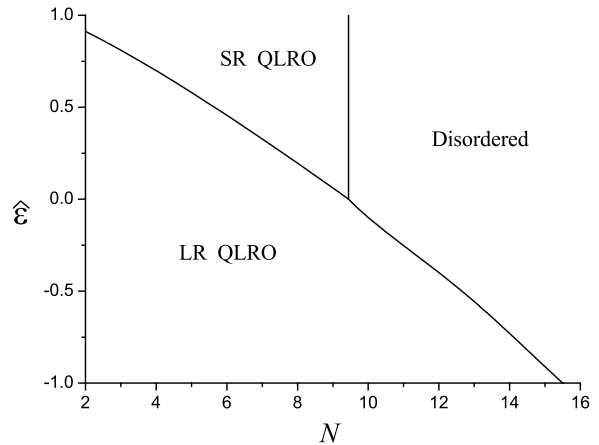}
\caption{Phase diagram of the RA model below the lower critical
dimension, $d<4+\sigma$ ($\varepsilon<\sigma$).} \label{fig3-ra}
\end{figure}

Below $d_{\mathrm{lc}}$, the LR RA model cannot develop a true long-range
order, but there can exist different types of QLRO. The SR QLRO phase is controlled by the
SR RA FP, with $r_1^*(\phi)=r_{\mathrm{SR}}^*(\phi)$ and  $r_2^*(\phi)=0$
computed in Ref.~\onlinecite{feldman00}. It was found to be stable
in the subspace of the SR correlated disorder for $N<N_c=9.4412$.
We find that it is also stable with respect to the LR correlated disorder for
$\sigma$ fulfilling inequality (\ref{ra-ineq}).
The new LR QLRO phase is controlled by an attractive
LR RA FP with $\pi$-periodic functions $r_1^*\ne0$ and $r_2^*\ne0$, which
we computed numerically by integrating
Eq.~(\ref{frg-1}) with the initial condition given by Eq.~(\ref{series-ra}) and
using $r_1^{*\prime\prime}(0)$ as a shooting parameter.
The exponents describing the correlation functions in the LR QLRO phase are given
by Eqs.~(\ref{exp-ra-1}) and (\ref{exp-ra-2}). Note that below $d_{\mathrm{lc}}$
we have $\eta>0$.
To check the stability of the LR RA FP, we substitute
it in Eq.~(\ref{frgl-2}) and obtain $y_2=N\cos^2 \phi-1$ and
\begin{equation}
  a_0=\frac{2\lambda N}{4r_1^{*\prime\prime}(0)-1}.
\end{equation}
The eigenfunction $y_1(\phi)$ and eigenvalue $\lambda$ are computed numerically
using shooting. This allows us to determine the stability regions for different
QLRO phases, which are shown in Fig.~\ref{fig3-ra}. As one can see from
Fig.~\ref{fig3-ra} in contrast to SR QLRO,  LR QLRO can exist even for
$4<d<4+\sigma$ and $N>N_c$.

The above results may be relevant for the behavior of $^3$He in aerogels.
It has recently been  observed in NMR experiments that the phase A of $^3$He confined
in aerogel exhibits two different types of magnetic behavior called
$c$ and $f$ states.\cite{dmitriev06} Depending on the cooling,
one can obtain either a nearly pure $c$  state or a mixed $f+c$ state,
which gives two overlapping lines
$c$ and $f$ in the transverse NMR spectrum.
Although the pure $f$ state has not been
observed, there is an evidence that the $f+c$ state is inhomogeneous and consists
of regions with two different magnetic orders $c$ and $f$.\cite{dmitriev06}
The order parameter of the $^3$He-A  can be parametrized by
the complex vector $\bm{\psi}$ and the real unit vector $\hat{\mathrm{\bf d}}$,
which characterize the orbital and  magnetic anisotropy, respectively.\cite{volovik-book}
Only the orbital part of the order parameter given by the real unit vector
$\hat{\mathrm{\bf l}}=(i/2)[\bm{\psi},\bm{\psi}^*]$ interacts with the aerogel
matrix, which can be treated as quenched random anisotropy disorder.  The spin part
$\hat{\mathrm{\bf d}}$  is not coupled directly to the disorder, but  there is
a weak spin-orbit (dipole) interaction between $\hat{\mathrm{\bf l}}$
and $\hat{\mathrm{\bf d}}$ which
generates the NMR frequency shift.
In Ref.~\onlinecite{volovik06}, the existence of the two states was interpreted in terms of
different ``random textures'' of the field $\hat{\mathrm{\bf l}}$ in the Larkin-Imry-Ma
state (QLRO phase in our notation).
The measured dependencies of the transverse NMR signal on the tipping angle
can be explained if one assumes that in the $c$ state
the vectors $\hat{\mathrm{\bf l}}$ and $\hat{\mathrm{\bf d}}$ are almost uncorrelated,
while in the $f$ state they are partially locked. This is possible if
in the $c$ state the characteristic length of the texture, i.e., the Larkin length,
is much smaller than the characteristic length of the dipole interaction,
$L\ll \xi_D$, while in the $f$ state they are of the same order.\cite{volovik06}
However, the nature of different random textures
exhibiting different characteristic length scales is not clear.
Alternatively, one can try to interpret the $f$ state as a network
of topological defects pinned by the aerogel.\cite{volovik06}
We argue that these two different states
may be the SR and LR QLROs found above in the LR RA model.
Indeed, aerogel is a porous medium formed by tangled silicon  strands which exhibit
a fractal mass distribution.  NMR and small-angle X-ray scattering (SAXS)
experiments give the mass-to-distance relation
$M\sim x^{d_f}$, with the fractal dimensions $d_f$ varying in the range of $1.4$ -- $2.4$
depending on the fabrication process. This scaling holds up to the fractal correlation
length, which can exceed 100 nm.\cite{devreux90}
Thus, the effective RA disorder is expected to
be long-range correlated with  $\sigma=d_f$
up to the scale of the fractal correlation length or even more.\cite{vascuez-03}
As a result, the SR QLRO phase is unstable to formation
of islands with LR QLRO (see Fig.~\ref{fig3-ra}).
To compute the Larkin length, one has to solve the flow equations (\ref{frg-1})
and (\ref{frg-2}) starting from a particular bare disorder correlator and
looking for the scale at which the cusp is developing. \cite{fedorenko06}
We can estimate the Larkin length using Flory-type arguments.\cite{feldman00} Assuming
that $J$ is an elastic constant and $R$ is a strength of disorder,
we obtain $L_{\mathrm{SR}} \sim (J/R)^{2/\varepsilon}$ and
$L_{\mathrm{LR}} \sim (J/R)^{2/(\varepsilon-\sigma)}$ for the SR and LR disorders,
respectively. For aerogel, we have $\varepsilon=1$ and $\sigma \approx 1.4\ldots 2.4$ so that the
Larkin lengths in both phases may differ significantly.
This can explain  the experimentally observed coexistence of regions with different
spin states.  Further investigations, however, are clearly needed.

\section{Summary}
\label{sec-final}

In this work, we investigated the long distance properties of the
$O(N)$ model with  random fields and random
anisotropies correlated as $1/x^{d-\sigma}$ for
large separation $x$. We
derived the functional renormalization group equations to one-loop
order, which allow us to describe the scaling behavior of the
models below and above the lower critical dimension
$d_{\mathrm{lc}}=4+\sigma$. Using a double $\varepsilon=d-4$ and
$\sigma$ expansion, we obtained the phase diagrams
and computed the critical exponents to first order in
$\varepsilon$ and $\sigma$. For the LR RF model, we found that the
truncated RG developed in Ref.~\onlinecite{chang84} to study the
critical behavior above the lower critical dimension is able to
give the correct one-loop values of exponents $\eta$ and
$\bar{\eta}$, but not the phase diagram  and
the critical exponent $\nu$ except for the region controlled by the weakly nonanalytic
LR TT FP.  Thus, although the truncated RG overcomes the
dimensional reduction, it fails to reproduce all properties which can be
obtained using the functional renormalization group.
We found a new LR QLRO phase
existing in the LR RF model below the lower critical dimension
for $N<3$ and determined the regions of its stability in the
$(\varepsilon,\sigma,N)$ parameter space. We obtained
that the weak LR correlated disorder does not change the critical behavior
of the RA model above $d_{\mathrm{lc}}$ for $N>N_c=9.4412$, but can create
a new LR QLRO phase below $d_{\mathrm{lc}}$. The existence of two QLRO phases
in LR RA systems may explain the two different states of $^3$He-A in aerogel
observed recently in NMR experiments.\cite{dmitriev06}
However, many questions are still open. In
particular,  the paramagnetic-ferromagnetic transition
should exist also for $N<N_c$, though it was not found in the one-loop
approximation. Even for the SR correlated disorder, it is still
unclear whether it remains perturbative.\cite{feldman02,doussal06}
It would also be interesting to find a connection with the
replica symmetry breaking picture.\cite{ledoussal03-3}

\begin{acknowledgments}
We would like to thank Pierre Le Doussal and Kay Wiese
for many stimulating discussions and critical reading of
the manuscript. We are also grateful to G.E. Volovik
for drawing our attention to recent results on superfluid helium
in aerogels. F.K. thanks LPT ENS  for hospitality during this work.
A.A.F. acknowledges support from the European Commission
under contract No. MIF1-CT-2005-021897.
\end{acknowledgments}

\end{document}